\documentclass[12pt]{article}
\usepackage[a4paper, top=16mm, text={170mm, 248mm}, includehead, includefoot, hmarginratio=1:1, heightrounded]{geometry}
\usepackage{amsmath,amssymb,mathrsfs,amsthm,tikz,shuffle,paralist}
\usepackage{dsfont}
\usepackage{color}

\usepackage{authblk}

\usepackage{hyperref}

\definecolor{darkred}{RGB}{173,34,48}
\hypersetup{bookmarksnumbered=true,
            colorlinks,
            linkcolor=darkred,
            citecolor=blue}

\theoremstyle{definition}
	\newtheorem{para}{}[section]
		
	\newtheorem{defi}[para]{Definition}
\theoremstyle{plain}

	\newtheorem{pro}[para]{Proposition}
	\newtheorem*{pro*}{Proposition}

\newcommand{\dif}{\mathrm{d}} 

\DeclareSymbolFont{lettersA}{U}{txmia}{m}{it}
\DeclareMathSymbol{\piup}{\mathord}{lettersA}{25}
\DeclareMathSymbol{\muup}{\mathord}{lettersA}{22}
\DeclareMathSymbol{\deltaup}{\mathord}{lettersA}{14}

\definecolor{shadecolor}{RGB}{32,32,32}
\definecolor{textcolor}{RGB}{204,204,255}

\date{\today}
\title{Blowing up Stringy Canonical Forms: An Algorithm to 
Win a Simplified Hironaka's Polyhedra Game}

\author[a,b]{Zhenjie Li}
\author[a,b]{Chi Zhang}
\affil[a]{\small CAS Key Laboratory of Theoretical Physics, Institute of Theoretical Physics, Chinese Academy of Sciences, Beijing 100190, China}
\affil[b]{\small School of Physical Sciences, University of Chinese Academy of Sciences, No.19A Yuquan Road, Beijing 100049, China}
\affil[ ]{Emails: \texttt{\{lizhenjie, zhangchi\}@itp.ac.cn}}

\begin{document}

\maketitle

\begin{abstract}
We provide an efficient method of blowing up to compute leading order contributions of the recently introduced stringy canonical forms. 
The method is related to the well-known Hironaka's polyhedra game, and the given algorithm is also useful on similar problems, \textit{e.g.} sector decomposition.
\end{abstract}
\newpage

\begin{center}
	\rule{1.0\textwidth}{1pt}
\end{center}
\tableofcontents

\bigskip
\begin{center}
	\rule{1.0\textwidth}{1pt}
\end{center}

\section{Introduction}

Very recently, a vast generalization of tree level strings integrals has been proposed in \cite{Arkani-Hamed:2019mrd}. This generalization is realized by identifying the Parke-Taylor form  
\[
	\mathsf{PT}(n):=\frac{\dif^{n}z}{\mathrm{SL}(2,\mathbb{R})} \prod_{i=1}^{n}\frac{1}{z_{i}-z_{i+1}}
\]
as the canonical form of the moduli space $\mathcal{M}_{0,n}^{+}$ and the Koba-Nielson factor as a regulator of the divergent integral $\int \mathsf{PT}(n)$. With a positive parameterization  $\mathbf{x}:= \{x_{1},\cdots, x_{n-3}\}$  of $\mathcal{M}_{0,n}^{+}$, $\mathsf{PT}(n)$ becomes the canonical form $\prod_{i=1}^{n-3} \dif \log x_{i}$ of $\mathbb{R}_{+}^{n-3}:=[0,\infty]^{n-3}$ and the Koba-Nielson factor $\prod_{i<j} (z_{i}-z_{j})^{s_{ij}}$ becomes a product of powers of some Laurent polynomials $p_{I}(\mathbf{x})$, then string integrals end up with the form of
\begin{equation} 
	\int_{\mathbb{R}_{+}^{D}} \prod_{i=1}^{D}\frac{\dif x_{i}}{x_{i}}x_{i}^{\alpha' X_{i}}\prod_{I}p_{I}(\mathbf{x})^{-\alpha'c_{I}}	\:, \label{int1}
\end{equation}
where $D=n{-}3$, $X_{i}$ and $c_{I}$ are linear combinations of Mandelstam variables $s_{ij}$'s. A \emph{stringy} canonical form (or a \emph{stringy} integral) is an integral of form eq.(\ref{int1}) but with arbitrary subtraction-free polynomials $p_{I}$.

In \cite{Arkani-Hamed:2019mrd}, many important properties of stringy integrals are well studied, especially the relation between their field theory limit, {\it{i.e}}. the limit of $\alpha'\to 0$, and the Minkowski sum $N_{P}$ of Newton polytopes of $\{p_I\}$.
More concretely, {\it{ i) the leading order of a stringy integral is given by the volume of the dual polytope, or the canonical function $\underline{\Omega}(N_{P})$, of $N_{P}$, and ii) a bijection between $\mathbb{R}_{+}^{D}$ and $N_{P}$ is given by the saddle point equation of the regulator, that is the so-called scattering-equation map $\Phi$,}}
\begin{equation} \label{SEmap}
	X_{i}= \sum_{I}c_{I}\frac{\partial \log p_{I}}{\partial \log x_{i}}\:, \qquad  \text{for }i=1,\cdots,D.
\end{equation}	
In general, however, it's difficult to calculate the leading order of stringy integrals directly from the above two properties.
On the one hand, performing Minkowski sum for polytopes analytically is nearly impossible. On the other hand, to obtain the canonical form $\Omega(N_{P})$ from the pushforward $\Phi_{\ast}\bigl(\bigwedge_{i=1}^{D} \dif \log x_{i}\bigr)$ requires to solve highly non-linear equations, just like the case in CHY formalism.

The main purpose of this article is to show an efficient method, {\it{blowing up}}, 
to calculate the leading order of the integral eq.\eqref{int1}  with respect to $\alpha'$. This is a general method which is closely related to the so-called \emph{Hironaka's polyhedra game}~\cite{hironaka1967} and also generally used in the calculation of Feynman diagrams to disentangle the singularities where this method is called \emph{sector decomposition} (see \textit{e.g.} \cite{Binoth:2000ps,Binoth:2003ak,Heinrich:2008si}). We will see that the situation is easier when this method is applied in stringy integrals due to some features of canonical forms, and especially its closed relation to the polytope $N_{P}$. This method is based on two simple observations: i) the leading order contribution in $\alpha'$-expansion of stringy integrals arises from each vertex of the integration region $\mathbb{R}_{+}^{D}$, ii) suppose that all $p_I(0)\neq 0$ in the integral eq.\eqref{int1}, then the integral at the neighbourhood of the origin becomes 
\begin{equation}\label{int2}
	\int_{[0,\epsilon]^D}\prod_{i=1}^D\frac{\dif x_i}{x_i}x_i^{\alpha' X_i} \prod_{I}p_{I}(0)^{-\alpha' c_{I}}
	=\prod_{i=1}^D\frac{\epsilon^{\alpha' X_i}p_{I}(0)^{-\alpha' c_{I}}}{\alpha' X_i}
	= \frac{1}{(\alpha')^D}\frac{1}{X_1\cdots X_D}+O(\alpha'^{-D+1}).
\end{equation}
Generally, it would not be such simple case, some polynomials $p_{I}$ may vanish or even is singular\footnote{A point $Q$ is a singular point of the surface defined by $p(x_{1},\dots,x_{n})=0$ if $\partial p(Q)/\partial x_{i} =0$ for all $i$~%
\cite{shafarevich1994basic}. } at the origin (or a vertex).
They can be overcome by a series of blows-ups, after which the stringy integral is decomposed into many integrals like eq.\eqref{int2}, then the leading order is given by a summation. %
Essentially, finding such a series of blow-ups is equivalent to give a winning strategy for the Hironaka's polyhedra game.
Besides, we can introduce an extra operation $x_i\mapsto x_i^c$ 
for any variable $x_i$ and any positive rational number $c$ 
which keeps the form of the integral eq.\eqref{int2} so that it leads to a simplified 
version of the Hironaka's polyhedra game. It's enough for us to win this 
simplified game to calculate the leading order of a stringy integral.

In section 2, we review the definition of blow-up and describe its relation to Hironaka's polyhedra game.
Section 3 uses several examples arising from cluster stringy integrals to illustrate this method.
In section 4, we introduce a new geometric viewpoint to approach a simplified version of Hironaka's polyhedra game and find a new algorithm to win.
Section 5 contains discussion and outlook. 

\section{Blow-up and Hironaka's Polyhedra Game}

In this section, we first use some heuristic examples 
to show the general procedure to calculate the leading order of stringy integrals
by blowing up and catch some important features of this calculation.
Then we go to the general case and show the equivalence of this procedure 
and the famous Hironaka's polyhedra game.

Before going into details of blowing up, let us clarify the problem mentioned in introduction. To this end, it is useful to recover a more general form of string integrals
\begin{equation*}
  \mathcal{I}(\alpha';W_{I})=\int_{\mathcal{P}} \Omega_{\mathcal{P}}(\mathbf{x}) \prod_{I}p_{I}(\mathbf{x})^{\alpha'W_{I}}\,,
\end{equation*}
where $\mathcal{P}$ is some positive geometry (in our case, it is some simple polytope) of dimension $D$ parameterized by $\mathbf{x}:= \{x_{1},\ldots,x_{D}\}$, $\Omega_{\mathcal{P}}$ is its associated canonical form~\cite{Arkani-Hamed:2017tmz}, and $p_{I}(\mathbf{x})$ are polynomials vanishing at boundaries of $\mathcal{P}$ and hence regulate the logarithm divergence of $\Omega_{\mathcal{P}}$ at boundaries. For $p_{I}(x)^{\alpha' W_{I}}$ to be single valued in $\mathcal{P}$, we require that $p_{I}(\mathbf{x})$ is nonnegative in the interior of $\mathcal{P}$. 

\begin{figure}[h]
\begin{center}
\includegraphics[scale=0.75]{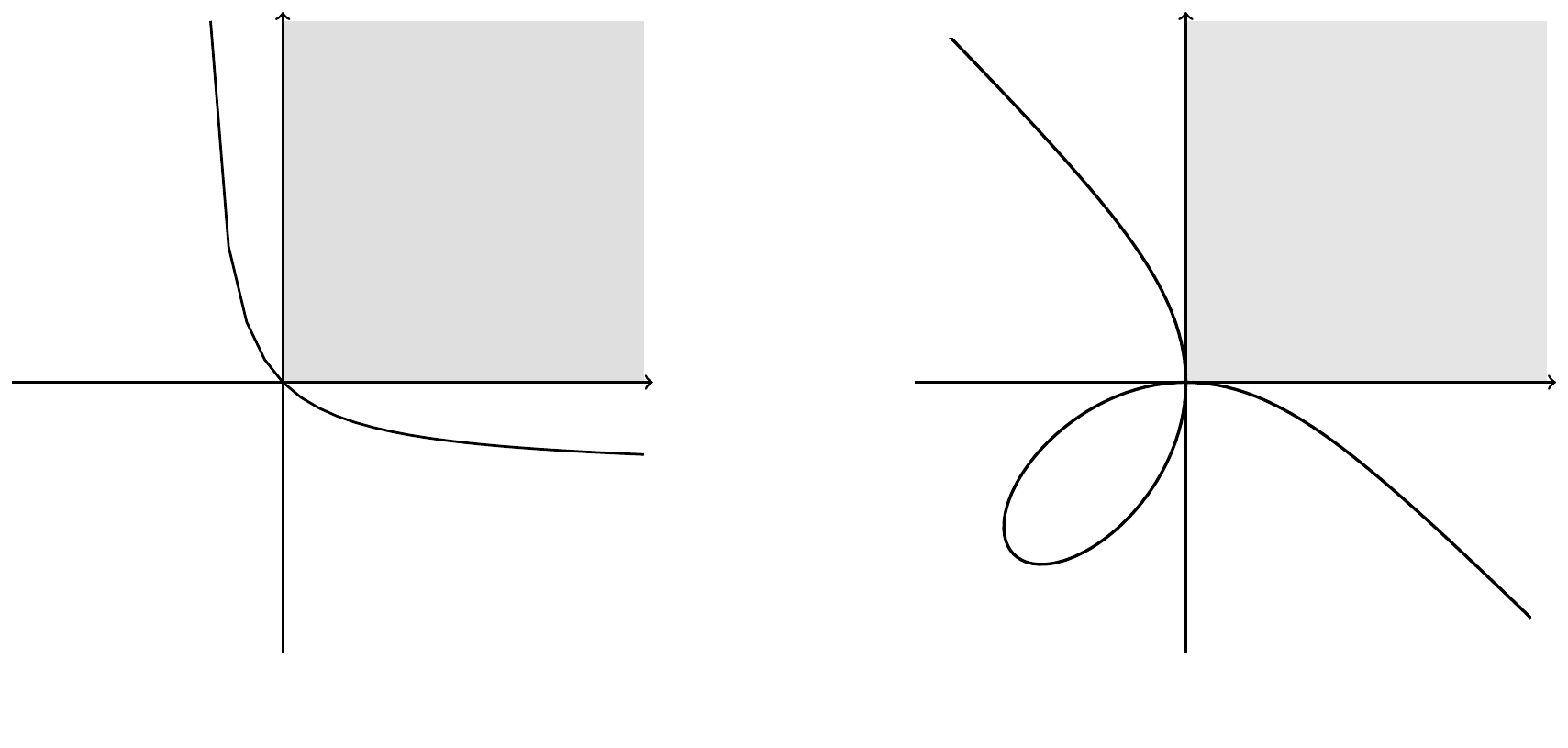}
\end{center}
\vspace{-5ex}
\caption{Left: $p=x+y+xy$, Right: $p=x^3+y^3+xy$}
\label{fig_1}
\end{figure}

The leading order contribution of $\mathcal{I}$ with respect to $\alpha'$ arises from the integral over the neighbourhood of each vertex of $\mathcal{P}$. For any vertex, the leading order contribution can be trivially obtained if there are just $D$ polynomials $p_{I}$ vanishing and regular (non-singular) at this vertex as in \eqref{int2}, and such vertex is called {\it{normal crossing}} since this vertex has been crossed $D$ times geometrically. All troubles are caused by vertices which are {\it not} normal crossing. For example (see Fig. \ref{fig_1}), consider the integral 
\begin{equation}\label{exam:1}
	\int_0^\infty \int_0^\infty \frac{\dif x}{x}\frac{\dif y}{y}x^{\alpha' X}
	y^{\alpha' Y}(x+y+ xy)^{-\alpha' c},
\end{equation}
the vertex $(0,0)$ is crossed by $x=0$, $y=0$ and $x+y+xy=0$ once. 
For the integral
\begin{equation}\label{exam:2}
\int_0^\infty \int_0^\infty \frac{\dif x}{x}\frac{\dif y}{y}x^{\alpha' X}
y^{\alpha' Y}(x^3+y^3+ xy)^{-\alpha' c},
\end{equation}
the vertex $(0,0)$ is crossed by $x=0$ and $y=0$ once and by $x^3+y^3+xy=0$ twice. 

The main tool used in this article to solve this problem is blowing up (or blow-up).
Briefly speaking, blow-up (along the original point) introduces extra dimensions for the original point such that
curves crossing it in different directions are lifted to new curves intersecting the extra dimensions in different points.

Let's first consider a simple but heuristic example of blowing up. 
Suppose there's a family of lines $\{l_i: a_ix+b_iy=0\}_{i=1,\dots,n}$ on the plane 
$\mathbb R^2$
crossing the original point $(0,0)$. 
If we want know how these lines cross the point $(0,0)$, we can introduce 
new variables $u$, $v$ and $t$ such that
\begin{equation} \label{2dimblowup}
	x=ut,\quad y=vt,
\end{equation} 
where $[u:v]$ is a projective coordinate, since any common factor of $u$ and $v$ can be absorbed into the definition of $t$, the line $l_i:a_ix+b_iy=0$ hence becomes a line parameterized 
by $t$ with an extra point $[-b_i:a_i]\in \mathbb P^1$. These extra points in $\mathbb P^1$ tell us how these line approach $(0,0)$ so that they can be used to distinguish different lines.
In other words, we replace the point $(0,0)$ with $\mathbb P^1$, which is the extra dimension mentioned above.

Now let's carefully consider the first example, the integral \eqref{exam:1}
\[
	I=\int_0^\infty \int_0^\infty\frac{dx}{x}\frac{dy}{y}x^{\alpha' a}y^{\alpha' b}(x+y+x y)^{-\alpha' c},
\]
whose leading contribution comes from the neighbourhoods of four vertices $(0,0)$, $(0,\infty)$, $(\infty,0)$ and $(\infty,\infty)$. One can easily see that the vertices $(0,\infty)$, $(\infty,0)$ and $(\infty,\infty)$ are normal crossing. The consequence of this fact is the integrand decouples into powers of $x$ and $y$, then the leading contribution can be trivially obtained, for example 
\begin{align*}
	I(0,\infty)&=\int_{0}^\epsilon\int_{1/\epsilon}^\infty \frac{dx}{x}\frac{dy}{y}x^{\alpha' a}y^{\alpha' b}(x+y+xy)^{-\alpha' c} \\
	&\approx \int_{0}^\epsilon\int_{1/\epsilon}^\infty \frac{dx}{x}\frac{dy}{y}x^{\alpha' a}y^{\alpha' (b-c)}= 
	\frac{1}{{\alpha'}^2}\frac{1}{a} \frac{1}{c-b}+O({\alpha'}^{-1}),
\end{align*}
where we throw away $x$ and $xy$ in $x+y+xy$ after $\approx$ because in this case $x\ll y$
so that they have no contribution to the leading order when $\alpha' \to 0$.
In this article, the symbol $\approx$ is used to relate two expressions 
(usually integrals) with the same leading terms 
or two polynomials which contribute the same leading terms in the integration.
Similarly, $I(\infty,0)\approx ({\alpha'}^2 b (c-a))^{-1}$ and $I(\infty,\infty)\approx (\alpha'^{2}(c-a) (c-b))^{-1}.$

The vertex $(0,0)$ is harder due to the behaviour of the mixed factor $(x+y+xy)^{-\alpha' c}$ at the neighbourhood of this vertex. 
However, we can always drop the term $xy$ which becomes irrelevant 
when $(x,y)\to (0,0)$ because $xy\ll x$, $y$. 
For remaining terms $x+y$, 
we blow up the plane along $(0,0)$ by introducing new variables 
$(t,[u:v])$ defined by equations $x=tu$ and $y=tv$ in
$\mathbb R\times\mathbb P^1$. 
Because $x$, $y\geq 0$ in the integral region,
the vertex $(0,0)$ is blown up to an $1$-simplex 
$\mathbb P^1_+=\{[u:v]\,:\,u,v\geq 0\}$ 
with two new vertices $u=0$ and $v=0$ under this map, 
and it is easy to see that the factor $(x+y)$ is dominated by $x$ 
near the vertex $u=0$ or by $y$ near the vertex $v = 0$. More precisely,
the $1$-simplex $\mathbb P^1_+$ can be identified with 
the unit interval $[0,1]$ by setting $u+v=1$, so
 \begin{equation}\label{canonicalformunderblowup}
	\frac{dxdy}{xy} = \frac{du}{u(1-u)}\frac{dt}{t},
 \end{equation}
and
\begin{equation*}
	I(0,0)\approx\int_0^\epsilon\frac{dt}{t} t^{\alpha'(a+b-c)}
	\int_0^{1}\frac{du}{u(1-u)}u^{\alpha' a}(1-u)^{\alpha' b}
	= \frac{1}{\alpha'}
	\frac{\epsilon^{\alpha'(a+b-c)}}{a+b-c}
	\int_0^{1}\frac{du}{u(1-u)}u^{\alpha' a}(1-u)^{\alpha' b},
\end{equation*}
the left one-dimensional integral of $u$ is again a stringy integral, and
its leading order comes from two vertices $u=0$ and $v=0$ ($u=1$). Therefore,
\[
	I(0,0)
	= \frac{1}{{\alpha'}^2} \frac{1}{a+b-c}
	\biggl(\frac 1a+\frac 1b\biggr) +O({\alpha'}^{-1}).
\]
\begin{center}
\begin{tikzpicture}[scale=0.8]
\fill[gray!30,opacity=0.4] (-6.9,1.4) -- (-3.9,1.4) -- (-3.9,4.4) -- (-6.9,4.4);
\draw[->] (-7,1.4) -- (-3.9,1.4);
\draw[->] (-6.9,1.3) -- (-6.9,4.4);
\draw[densely dotted]  (-6.9,1.4) ellipse (0.5 and 0.5);
\draw[->] (-6.9,1.2) .. controls (-6.9,0.4) and (-6.7,0) .. (-4.7,0);
\fill[gray!30,opacity=0.4] (-3.5,-0.8) -- (-1.5,-0.8) -- (-1.5,2.2) -- (-4.5,2.2) -- (-4.5,0.2);
\draw[->] (-4.6,-0.8) -- (-1.5,-0.8);
\draw[->] (-4.5,-0.9) -- (-4.5,2.2);
\draw (-3.5,-0.8) -- (-4.5,0.2);
\draw[densely dashed,->] (-4.5,0.2) -- (-3.2,-1.1)node[right]{$u$};
\draw[-]  plot[smooth, tension=.7] coordinates {(-4.4,0.3) (-2.6,1.7) (-0.8,2)};
\draw[->] (-0.8,2)--(-0.6,2);
\draw[-]  plot[smooth, tension=.7] coordinates {(-3.5,-0.9) (-2.4,-2.3) (-0.8,-2.6)};
\draw[->] (-0.8,-2.6)--(-0.6,-2.6);
\fill[gray!30,opacity=0.4] (0.8,-3.4) -- (2.8,-3.4) -- (2.8,-0.4) -- (-0.2,-0.4) -- (-0.2,-3.4);
\draw[->] (-0.3,-3.4) -- (2.8,-3.4);
\draw[->] (-0.2,-3.5) -- (-0.2,-0.4);
\fill[gray!30,opacity=0.4] (0.8,1) -- (2.8,1) -- (2.8,4) -- (-0.2,4) -- (-0.2,1);
\draw[->] (-0.3,1) -- (2.8,1);
\draw[->] (-0.2,0.9) -- (-0.2,4);
\node at (-2.8,2) {\small $u\to 0$};
\node at (-2.0,-2) {\small $v\to 0$};
\node at (-4,-1) {\small $t$};
\node at (-4.6,-0.4) {\small $t$};
\node at (2.8,0.8) {\small $u$};
\node at (-0.4,4) {\small $t$};
\node at (-0.4,-0.4) {\small $t$};
\node at (2.8,-3.6) {\small $v$};
\end{tikzpicture}
\end{center}
Finally,
\begin{align*}
	I&\approx I(0,0)+I(0,\infty)+I(\infty,0)+I(\infty,\infty)\\
	 &\approx \frac{1}{{\alpha'}^2}
	 \biggl(
\frac{1}{b (a+b-c)}+\frac{1}{b (c-a)}+\frac{1}{a (c-b)}+\frac{1}{(c-a) (c-b)}+\frac{1}{a (a+b-c)}
	 \biggr).
\end{align*}

From the last example, we see that polynomials are not canonical objects 
for the leading order of a integral because different polynomials may give the 
same contribution. For example, we can throw away irrelevant terms in the 
polynomial. Besides, the integral also doesn't depend on positive coefficients
in the polynomial which we will explain later.
A natural question thus arises. What's the canonical object to describe the 
leading contribution of a polynomial in an integral near a vertex? The answer is 
the \textit{Newton polyhedron} of this polynomial.

\begin{defi}[Newton polyhedron]
Let $p=\sum_{I}s_I x^{n^I}$ be a polynomial with positive coefficients.
For each term $s_I x^{n^I}$, we assign a cone 
$C_{I}=(n^I+\mathbb R^D_{+}):=\{(n^I_1+v_1,\dots,n^I_D+v_D)\,:\, 
\text{$v_i\geq 0$ for all $i$}\}$. The 
\textit{Newton polyhedron} $C[p]$ of the polynomial $p$ 
is the \textit{convex hull} of cones $\{C_I\}$, 
\textit{i.e.} the smallest convex set contains these cones.
\end{defi}




\begin{pro}\label{pro:1}
The leading order of integrals
\[
	I=\int_{[0,\epsilon]^D} \left(\prod_{i=1}^D\frac{\dif x_i}{x_i}x_i^{\alpha' X_i}\right)
	p^{-\alpha' c} \prod_a q_a^{-\alpha' c_I}
\]
only depends on the Newton polyhedron $C[p]$.
\end{pro}

From the theorem in \cite{Arkani-Hamed:2019mrd}, the leading order of the integral eq.\eqref{int1} 
only depends on the Minkowski sum of Newton polytopes of polynomials $p_I$. 
For a polynomial $q=\sum_I a_I x^{n^I}$, the {\textit{Newton polytope}} is the convex hull of vectors 
$\{n^I=(n^I_1,\dots,n^I_D)\}$, while \textit{Minkowski sum} (or vector sum) of two set $A$ and $B$ is the set $\{x+y\,:\,x\in A,\,y\in B\}$. Note that the Newton polyhedron of a polynomial is just the Minkowski sum of Newton polytope and $\mathbb R_+^D$.

Therefore, the above proposition is like the local version of this theorem, 
so it may seem trivial. However, it's more convenient to use the local version because 
the Newton polyhedron is not as rigid as the Newton polytope, 
and it just looks like a `corner' of the Newton polytope.


By using the language of the polyhedra for polynomials, the goal of blowing up is clear.
If one gets the polyhedron of a polynomial $p$ like an {\it{orthant}}
\begin{center}
\begin{tikzpicture}[scale=0.75,baseline={([yshift=-.5ex]current bounding box.center)}]
\fill[gray!20] (2,3) -- (0,3) -- (0,0) -- (3,0) -- (3,1) -- (3,3);
\draw[->](-1.1,-0.5) -- (3,-0.5);
\draw[->](-1,-0.6) -- (-1,3);
\draw[->] (-1,-0.5) -- (0,0);
\node at (-0.5,0) {$v$};
\end{tikzpicture}
\end{center}
then we call that $p\approx x^v$ is \emph{decoupled}. In other words, $p$ 
has the form of $p(x)=x^v(a+q(x))$ for a positive constant $a$, a vector $v$ and 
polynomial $q$ with $q(0)=0$. Suppose $p=x^v(a+q(x))$ is a decoupled polynomial, then
the integral
\[
	I=\int_{[0,\epsilon]^D} \left(\prod_{i=1}^D\frac{\dif x_i}{x_i}x_i^{\alpha' X_i}\right)
	p^{-\alpha' c} \prod_I q_I^{-\alpha' c_I}
	\approx\int_{[0,\epsilon]^D} \left(\prod_{i=1}^D\frac{\dif x_i}{x_i}x_i^{\alpha' (X_i-cv_i)}\right)
	\prod_I q_I^{-\alpha' c_I}
\] 
is reduced to a new integral with fewer polynomials but with the same leading order. 
This is the simplest case of Proposition \ref{pro:1}, and it's very easy to prove.
In fact, other terms in the cone $C[p]$ go to zero faster than $a x^v$,
and the factor $a^{-\alpha'c}=1+O(\alpha')$ introduced by the coefficient $a$ 
doesn't affect the leading order of the integral.

Therefore, our aim is to find a series of blow-ups to make $p$ decoupled at 
all generated vertices.
If we can find it, the proof of Proposition \ref{pro:1} will be reduced to the trivial 
and proven case where $p$ is decoupled.
Since blow-ups never increase the number of polynomials in the integral, 
we only need to consider integrals that only contain one 
polynomial
\[
	I=\int_{[0,\epsilon]^D} \left(\prod_{i=1}^D\frac{\dif x_i}{x_i}x_i^{\alpha' X_i}\right)
	p^{-\alpha' c}.
\] 

Now it's a good time to consider general blow-ups and see their effect on integrals and polyhedra. 

The above blow-up \eqref{2dimblowup} is called the blow-up of the plane along the original point. In this article, we will consider the general blow-up of $\mathbb R^D$ along subspace $\mathbb R^n$ defined by $x_{i_1}=\cdots=x_{i_n}=0$. For those boundaries defined by %
$x_{i}=\infty$, we can change the variables by $x_i\mapsto 1/x_i$.
The blow-up is the variety in $\mathbb R^{D}\times \mathbb P^{n-1}$ defined by 
\begin{equation}
	\{x_{i_j}y_k=x_{i_k}y_j\,:\, 1\leq j<k\leq n\}\:,  \label{genblowup}
\end{equation}
where $[y_1:\dots:y_n]$ is the projective coordinate. These equations can be easily solved by $x_{i_j}=ty_{j}$ with
$\sum_{i=1}^n y_i=1$ for $t\neq 0$. 
The integral near the boundary defined by $x_{i_1}=\cdots=x_{i_n}=0$ becomes the integral over an interval of $t$ and a $(n-1)$-simplex $\mathbb P_+^{n-1}$ 
defined by
\[
	\sum_{i=1}^n y_i=1,\quad \text{$y_i>0$ for all $1\leq i\leq n$}.
\]
This blow-up produces $n$ new vertices 
$
	\{v_i:x'=0,t=0,y_i=1\}
$, where $x'$ is the other coordinates of $\mathbb R^D$.
Near the new vertex defined by $y_k=1$, it's equivalent to do 
the following change of variables
\[
	x'\to x',\quad x_{i_j}\to ty_j\quad \text{for $j\neq k$},\quad x_{i_k}\to t,
\]
or by reusing the name of $x_{i_k}$ for $t$ and $x_{i_j}$ 
for $y_j$ to save the namespace,
\begin{equation}\label{sec_doc}
	x_{i_j}\to x_{i_k}x_{i_j}\quad \text{for $j\neq k$} ,
\end{equation}
which is related to the sector decomposition~\cite{Heinrich:2008si}. 
The neighbourhood of this vertex in the integral region 
can be taken as $0<x_i<\epsilon$ for all $i$.

As hinted by eq.\eqref{canonicalformunderblowup}, one important feature of $\Omega_{\mathcal P}$ is the `invariance' under blowing up, 
since the residue of the canonical form on the boundary is
the canonical form of the boundary.
Suppose we blow up the boundary defined locally by $x_1=\cdots=x_n=0$ in integral eq.\eqref{int1}. 
Let $x_i=ty_i$, where $[y_1:\dots:y_n]$ are positive projective coordinates. 
Near the boundary, the canonical form of $\Omega$ behaves as 
\[
	\Omega=\frac{dx_1}{x_1}\wedge \cdots\wedge\frac{dx_n}{x_n}
	\wedge \Phi(x'),
\]
where $x'$ are other coordinates. 
At each generated vertex of $(n-1)$-simplex $\mathbb P_+^{n-1}$, the canonical
form $\Omega$ is invariant under the change of variables eq.\eqref{sec_doc}.

Now let us consider the effect of blow-up on the polynomial $p$. 
For the blow-up of $\{x_{i_1},\dots,x_{i_n}\}$, the polynomial $p=\sum_I a_I x^{n^I}$ becomes 
$p'=\sum_I a_I x^{(n^I)'}$ by $x_{i_j}\mapsto x_{i_k}x_{i_j}$ for all $j\neq k$ in the neighbourhood of the $k$-th vertex, where
\[
	(n^I)'_{i_k}\longrightarrow \sum_{j=1}^n n^I_{i_j}.
\]
This can be visualized by the polyhedron. For instance, the Newton polyhedron of $p(x)=x^3+xy+y^3$ is the gray polyhedron, we first get the green polyhedron by $ y\mapsto xy$, but it's not the wanted (decoupled) polyhedron, so we blow up it by $y\mapsto xy$ again, then we get the red polyhedron.
\begin{center}
\begin{tikzpicture}[scale=0.75]
	\fill[gray!20] (0,4) -- (0,3) -- (1,1) -- (3,0) -- (4,0) -- (7.5,4);
	\fill[green!20] (2,4) -- (2,1) -- (3,0) -- (4,0) -- (7.5,4) --cycle;
	\fill[red!20] (3,4) -- (7.5,4) -- (7.5,0) -- (3,0)-- cycle;
	\draw[thick] (0,4) -- (0,3) -- (1,1) -- (3,0) -- (7.5,0);
	\draw[->](-0.1,0) -- (7.6,0);
	\draw[->](0,-0.1) -- (0,4.1);
	\node[inner sep=1.5pt,circle,fill=blue] at (0,3) {};
	\node[inner sep=1.5pt,circle,fill=blue] at (1,1) {};
	\node[inner sep=1.5pt,circle,fill=blue] at (3,0) {};
	\node[inner sep=1.5pt,circle,fill=blue] at (2,1) {};
	\node[inner sep=1.5pt,circle,fill=blue] at (3,3) {};
	\node[inner sep=1.5pt,circle,fill=blue] at (3,1) {};
	\node[inner sep=1.5pt,circle,fill=blue] at (6,3) {};
	\draw[dashed,->] (0.2,3) -- (2.8,3);
	\draw[dashed,->] (1.2,1) -- (1.8,1);
	\draw[dashed,->] (3.2,3) -- (5.8,3);
	\draw[dashed,->] (2.2,1) -- (2.8,1);
	\node at (7.3,-0.2) {$x$};
	\node at (-0.2,3.8) {$y$};
\end{tikzpicture}
\end{center}
Note that we still need to deal with other vertices generated by blow-ups. 

Due to the behaviors of the canonical form and the polynomial $p$ under the blow-up. Finding a {\it{finite}} series of blow-ups such that the polynomial $p$ decouple in each generated vertices is essentially reformulated into a game by Hironaka~\cite{hironaka1967}. In this game, according to the given polynomial $p$, two players $\mathcal{P}_{1}$ and $\mathcal{P}_{2}$ make the following moves:
\begin{enumerate}
	\item $\mathcal{P}_{1}$ choose a set of variables $\{x_{i_1},\dots,x_{i_n}\}$.
	\item $\mathcal{P}_{2}$ choose one variable $x_{i_k}$ out of them and make variable substitutions  $x_{i_j} \to x_{i_k}x_{i_j}$ for $j\neq k$
\end{enumerate}
If $p$ becomes decoupled (or geometrically, $C[p]$ becomes an orthant), then player $\mathcal{P}_1$ wins, otherwise they start a new round by using the new generated polynomial. If this never occurs, player $\mathcal{P}_{2}$ will have won.
For example, 
\[
	p(x_1,x_2,x_3)=x_1x_3^2+x_2^2+x_2x_3,
\]
if $\mathcal{P}_{1}$ choose $\{x_1,x_2\}$, $\mathcal{P}_{2}$ choose $x_{1}$ and make the variable substitution $x_2\mapsto x_1x_2$, 
since under this variable substitution
\[
	p(x_1,x_2,x_3)\mapsto x_1(x_1x_2^2+x_3^2+x_2x_3)=x_1 p(x_1,x_3,x_2),
\]
then $\mathcal{P}_{2}$ will win if $\mathcal{P}_{1}$ always choose such variables to blow up.




An winning strategy for $\mathcal{P}_{1}$ will tell us how to blow up the polynomial $p$ and hence calculate stringy integrals. There're many known winning strategies \cite{spivakovsky1983solution,zeillinger2005polyederspiele,zeillinger2006short,hauser2003hironaka},
and they have been used in many programs (see \textit{e.g.} \cite{Smirnov:2008py,Ueda:2009xx}).
In this article, we will give a new algorithm to win a simplified version of this game 
from a geometric viewpoint. The simplified version allows $\mathcal P_1$ to 
use another operation $x_j\mapsto x_j^c$ for any $j$ in the choosen set and any
positive rational number $c$, and the algorithm also tells us how to calculate 
the leading order of stringy integrals.
Before elaborating on this algorithm, we first give several simple examples to get a feeling of this blow-up method.

\section{Application and Example}

In this section, we will give several examples to illustrate this blow-up method, all examples come from the so-called cluster stringy integrals~\cite{Arkani-Hamed:2019mrd,Arkani-Hamed:2019plo} which are closely related to the origin string integrals. The dimension of all examples in this section is 2 or 3 and the regulating polynomials are simple, so the blow-up prescriptions can be designed \emph{ad hoc} without a universal algorithm.

\subsubsection*{Cluster string integral: $A_{2}$}
This integral is equivalent to the $Z$-integral $Z_{12345}(1,2,3,4,5)$~\cite{Mafra:2016mcc}, which in a positive parameterization takes the form of 
\[
	\mathcal{I}_{A_{2}}=\int_{\mathbb R_+^2} \frac{\dif x_1}{x_1}\frac{\dif x_2}{x_2}x_1^{\alpha' X}x_2^{\alpha' Y}
	(1+x_1)^{-\alpha' a}(1+x_2)^{-\alpha' b}(1+x_1+x_1x_2)^{-\alpha' c}
\] 
where $1+x_{1}$, $1+x_{2}$ and $1+x_{1}+x_{1}x_{2}$ are $F$-polynomials with the initial seed $A_{2}$ quiver %
(see \cite{fomin2007cluster} for cluster algebra). With variable substitutions
\[
	x_{1}=\frac{1-z_{3}}{z_{3}}\,,\qquad x_{2}=\frac{z_{3}-z_{2}}{z_{2}(1-z_{3})}
\]
and 
\[
	X=s_{12},\quad Y=s_{45},\quad a=-s_{24},\quad b=-s_{35},\quad c=-s_{25} \:,
	\] 
the integral $\mathcal{I}_{A_{2}}$ becomes $Z_{12345}(12345)$ under the usual gauge fixing $\{z_{1},z_{4},z_{5}\}\to\{0,1,\infty\}$. It is easy to see that vertices $(0,0)$, $(0,\infty)$ and $(\infty,0)$ are all normal crossing, then no blow up is needed and their contributions to leading order simply are
\begin{align*}
	&\mathcal{I}_{A_{2}}(0,0)=\frac{1}{\alpha'^{2}}\frac{1}{XY}+O(\alpha'^{-1}) \:, \\
	&\mathcal{I}_{A_{2}}(\infty,0)=\frac{1}{\alpha'^{2}}\frac{1}{(a+c-X)Y}+O(\alpha'^{-1}) \:, \\
	&\mathcal{I}_{A_{2}}(\infty,\infty)=\frac{1}{\alpha'^{2}}\frac{1}{(a+c-X)(b+c-Y)}+O(\alpha'^{-1}) \:.
\end{align*}
The vertex $(0,\infty)$ is not normal crossing, we bring this vertex to the origin by taking $x_{2}\to x_{2}^{-1}$, then the integral over the neighbourhood of this vertex reads
\[
    \mathcal{I}_{A_{2}}(0,\infty) \approx \int_{[0,\epsilon]^{2}} \frac{\dif x_{1}\,\dif x_{2}}{x_{1} x_{2}} x_{1}^{\alpha'X}x_{2}^{\alpha'(b+c-Y)}(x_{1}+x_{2}+x_{1}x_{2})^{-\alpha'c}
\]
where some irrelevant power functions of the form $(1+\cdots)$ have been dropped. This is exactly the example we used in section 2, then the leading order contribution of $\mathcal{I}_{A_{2}}$ is simply
\begin{align*}
   \mathcal{I}_{A_2}&=\frac{1}{\alpha'^{-2}}\biggl(\frac{1}{(a+c-X) (b+c-Y)}+\frac{1}{X (b+X-Y)} \\
   &\quad +\frac{1}{(b+c-Y) (b+X-Y)}+\frac{1}{Y (a+c-X)}+\frac{1}{X Y}\biggr) + O(\alpha'^{-1}) .
\end{align*}

\subsubsection*{Cluster stringy integral: $A_3$}

This integral is equivalent to the $Z$-integral $Z_{123456}(1,2,3,4,5,6)$, which in a positive parameterization takes the form of 
\begin{align*}
	\int_{\mathbb R_+^3} \frac{\dif x_1}{x_1}\frac{\dif x_2}{x_2}\frac{\dif x_3}{x_3}
	x_1^{\alpha' X}x_2^{\alpha' Y}x_3^{\alpha' Z}
	&(x_1 + 1)^{-\alpha' a_1}
	(x_2 + 1)^{-\alpha'a_2}
	(x_3 + 1)^{-\alpha'a_3}
	(x_1x_2 + x_1 + 1)^{-\alpha'a_4}\\
	&(x_2x_3 + x_3 + 1)^{-\alpha'a_5}
	(x_1x_2x_3 + x_1x_3 + x_1 + x_3 + 1)^{-\alpha'a_6}
\end{align*}
where 6 polynomials with constant term 1 are $F$-polynomials with the initial seed $A_{3}$ quiver. With variable substitutions
\[
x_{1}=\frac{(z_{2}-z_{3})(1-z_{4})}{(1-z_{2})(z_{3}-z_{4})}\:,\quad x_{2}=\frac{z_{4}-z_{2}}{z_{2}(1-z_{4})}\:, \quad
x_{3}=\frac{z_{2}}{1-z_{2}}
\]
and
\begin{gather*}
	X=s_{12},\quad Y=s_{123},\quad Z=s_{45} \\
	a_{1}=-s_{24},\quad a_{2}=-s_{36},\quad a_{3}=-s_{15},\quad a_{4}=-s_{26},\quad a_{5}=-s_{35},\quad a_{6}=-s_{25}	
\end{gather*}
the integral $\mathcal{I}_{A_{3}}$ becomes $Z_{123456}(1,2,3,4,5,6)$ under the usual gauge fixing $\{z_{1},z_{5},z_{6}\}\to\{0,1,\infty\}$. Of the 8 vertices, $(0,0,0)$, $(0,0,\infty)$, $(\infty,0,0)$, $(\infty,0,\infty)$ and $(\infty,\infty,\infty)$ are normal crossing ones, and their contributions to leading order is 
\begin{align*}
	\mathcal{I}_{A_{3}}^{(1)}&=\frac{1}{\alpha'^{2}}\biggl(\frac{1}{XYZ}+\frac{1}{F_{1}XY} +\frac{1}{F_{2}YZ}+\frac{1}{F_{1}F_{2}Y} 
	+\frac{1}{F_{1}F_{2}F_{3}}\biggr) +O(\alpha'^{-1})
\end{align*}	
where
\begin{align*}
	F_{1} &= a_{3}+a_{5}+a_{6}-Z \\
	F_{2} &= a_{1}+a_{4}+a_{6}-X \\
	F_{3} &= a_{2}+a_{4}+a_{5}+a_{6}-Y
\end{align*}
For the remaining vertices, we bring them to the origin by setting $x_{i}\to x_{i}^{-1}$,
\begin{align*}
	\mathcal{I}_{A_{3}}(0,\infty,0)&\approx\int_{[0,\epsilon]^{3}}\prod_{i=1}^{3}\frac{\dif x_{i}}{x_{i}} x_{1}^{\alpha' X}
	x_{2}^{\alpha' F_{3}}x_{3}^{\alpha' Z}(x_{1}+x_{2})^{-\alpha' a_{4}}(x_{2}+x_{3})^{-\alpha' a_{5}}
	\biggl(x_{2}+\sum_{i<j} x_{i}x_{j}\biggr)^{-\alpha' a_{6}} \\
	\mathcal{I}_{A_{3}}(\infty,\infty,0)&\approx \int_{[0,\epsilon]^{3}}\frac{\dif x_{i}}{x_{i}}x_{1}^{\alpha' F_{2}}x_{2}^{\alpha' F_{3}}x_{3}^{\alpha'Z} (x_{2}+x_{3})^{-\alpha'(a_{5}+a_{6})} \\
	\mathcal{I}_{A_{3}}(0,\infty,\infty)&\approx \int_{[0,\epsilon]^{3}}\frac{\dif x_{i}}{x_{i}}x_{1}^{\alpha' X}x_{2}^{\alpha' F_{3}}x_{3}^{\alpha' F_{1}} (x_{1}+x_{2})^{-\alpha'(a_{4}+a_{6})}
\end{align*}
where we have dropped some irrelevant terms according to our algorithm. The last two again are the cases we have encountered before, then a further attention is needed only for the first one, which can be decomposed into six normal crossing pieces 
\begin{equation*}
	\mathcal{I}_{A_{3}}(0,\infty,0)\approx
	\begin{cases}
		\int_{[0,\epsilon]^{3}}\prod_{i=1}^{3}\frac{\dif y_{i}}{y_{i}} y_{1}^{\alpha'X} y_{2}^{\alpha' F_{4}}
		y_{3}^{\alpha'F_{5} }  & \{x_{1}=y_{1}y_{2}y_{3}, x_{2}=y_{2}y_{3},x_{3}=y_{3}\} \\
		\int_{[0,\epsilon]^{3}}\prod_{i=1}^{3}\frac{\dif y_{i}}{y_{i}} y_{1}^{\alpha'X} y_{2}^{\alpha' (X+Z)}
		y_{3}^{\alpha'F_{5} }  & \{x_{1}=y_{1}y_{2}y_{3}, x_{3}=y_{2}y_{3},x_{2}=y_{3}\}  \\
		\int_{[0,\epsilon]^{3}}\prod_{i=1}^{3}\frac{\dif y_{i}}{y_{i}} y_{1}^{\alpha' F_{3}} y_{2}^{\alpha' F_{4}}
		y_{3}^{\alpha'F_{5} }  & \{x_{2}=y_{1}y_{2}y_{3}, x_{1}=y_{2}y_{3},x_{3}=y_{3}\}  \\
		\int_{[0,\epsilon]^{3}}\prod_{i=1}^{3}\frac{\dif y_{i}}{y_{i}} y_{1}^{\alpha' F_{3}} y_{2}^{\alpha' F_{6}}
		y_{3}^{\alpha'F_{5} }  & \{x_{2}=y_{1}y_{2}y_{3}, x_{3}=y_{2}y_{3},x_{1}=y_{3}\}  \\
		\int_{[0,\epsilon]^{3}}\prod_{i=1}^{3}\frac{\dif y_{i}}{y_{i}} y_{1}^{\alpha' Z} y_{2}^{\alpha' (X+Z)}
		y_{3}^{\alpha'F_{5} }  & \{x_{3}=y_{1}y_{2}y_{3}, x_{1}=y_{2}y_{3},x_{2}=y_{3}\}  \\
		\int_{[0,\epsilon]^{3}}\prod_{i=1}^{3}\frac{\dif y_{i}}{y_{i}} y_{1}^{\alpha' Z} y_{2}^{\alpha' F_{6}}
		y_{3}^{\alpha'F_{5} }  & \{x_{3}=y_{1}y_{2}y_{3}, x_{2}=y_{2}y_{3},x_{1}=y_{3}\}  
	\end{cases}
\end{equation*}
where each piece correspond to a simplex $0<x_{i}<x_{j}<x_{k}$, irrelevant terms are dropped again, and we have introduced
\begin{align*}
	&F_{4}=X+F_{3}-a_{4}-a_{6} \:, \\
	&F_{5}=F_{3}+X+Z-a_{4}-a_{5}-a_{6} \:, \\
	&F_{6}=F_{3}+Z-a_{5}-a_{6} \:.
\end{align*}
It can be easily checked that all $F$'s are planar and remaining poles, although the spurious pole $X+Z$ appears in the process, and the result for leading order is exactly the 6 point amplitude for bi-adjoint $\phi^{3}$ theory.

\subsubsection*{Cluster stringy integral: $C_2$}

This integral is equivalent to the $Z$-integral $Z(1^{+}2^{+})$ defined on the moduli space of paired punctures~\cite{Li:2018mnq}, which in a positive parameterization is 
\begin{align*}
\int_{\mathbb R_+^2} \frac{\dif x_1}{x_1}\frac{\dif x_2}{x_2}x_1^{\alpha' X}x_2^{\alpha' Y}
&(x_1 + 1)^{-\alpha' a} 
(x_2 + 1)^{-\alpha' b}\\
&(x_1x_2 + x_1 + 1)^{-\alpha' c}
(x_1^2x_2 + x_1^2 + 2x_1 + 1)^{-\alpha' d} 
\end{align*}
where 4 polynomials with constant term 1 are $F$-polynomials with a $C_{2}$ quiver as the initial seed.
This integral is related to $Z(1^{+}2^{+})$ (with gauge fixing $z_{0}=1$) by 
\begin{equation*}
	x_{1}=\frac{(1-z_{1})(z_{2}-z_{1})}{2z_{1}(1+z_{2})} \:,\qquad x_{2}=-\frac{(1+z_{1})^{2}}{(1-z_{1})^{2}}
\end{equation*}
and
\begin{align*}
	X=2s_{01}\:,\quad Y=s_{012}\:,\quad a=-2s_{1\tilde{2}}\:,\quad b=-s_{2\tilde{2}}\:,\quad 
	c=-2s_{02}\:,\quad d=-s_{0\tilde{0}}
\end{align*}
with the Mandelstam variables defined therein. In this case, only vertex $(0,\infty)$ is not normal crossing, we transform this vertex to the origin as before, then we have
\begin{align*}
	\mathcal{I}_{C_{2}}(0,\infty)\approx \int_{[0,\epsilon]^{2}} \frac{\dif x_{1}\dif x_{2}}{x_{1}x_{2}} x_{1}^{\alpha' X}x_{2}^{\alpha' G_{1}}
	(x_{1}+x_{2})^{-\alpha' c} (x_{1}^{2}+x_{2})^{-\alpha' d}
\end{align*}
where we have dropped irrelevant terms and introduced $G_{1}=b+c+d-Y$. This vertex can be decomposed into 3 normal crossing pieces by the following blow ups
\begin{align*}
	\mathcal{I}_{C_{2}}(0,\infty)\approx\left\{
	\begin{aligned}
		&\int_{0<x_{1}<x_{2}}\frac{\dif y_{1}\dif y_{2}}{y_{1}y_{2}} y_{1}^{\alpha' X}y_{2}^{\alpha' G_{2}} \\
		&\int_{0<x_{2}<x_{1}}\frac{\dif y_{1}\dif y_{2}}{y_{1}y_{2}} y_{1}^{\alpha' G_{1}}y_{2}^{\alpha' G_{2}}(y_{1}+y_{2})^{-\alpha'd} \approx
		\left\{\begin{aligned}
			\int_{0<y_{1}<y_{2}}\frac{\dif z_{1} \dif z_{2}}{z_{1}z_{2}}z_{1}^{\alpha' G_{1}}
			z_{2}^{\alpha'(G_{1}+G_{2}-d)} \\
			\int_{0<y_{2}<y_{1}}\frac{\dif z_{1}\dif z_{2}}{z_{1}z_{2}} z_{2}^{\alpha' G_{2}}
			z_{1}^{\alpha'(G_{1}+G_{2}-d)}
		\end{aligned} \right. 
	\end{aligned} \right. ,
\end{align*} 
where $G_{2}=G_{1}+X-c-d$, and the integral region for integration variables should be understood as $[0,\epsilon]^{2}$ while the subscripts of integrals indicate the integration region before blow-ups.

\section{Algorithm}\label{s:alg}

In this section, we design a new algorithm to `win' the Hironaka's
polyhedra game from a new viewpoint. The algorithm use not only blow-ups
but also an extra kind of operation, rescaling exponents of variables. 
If we change the variables by
\(
	x_i\mapsto x_i^{1/c_i}
\)
for $i\in S$, where $c_i$ is a positive rational number, then integral becomes
\[
	\prod_{i\in S}c_i\int
	\prod_{j\not\in S}\frac{\dif x_j}{x_j}x_j^{\alpha' X_j}
	\prod_{i\in S}\frac{\dif x_i}{x_i}x_i^{\alpha' c_iX_i}
	p(\{x_i^{c_i}\,:\, i\in S\};\{x_j\,:\, j\not\in S\})^{-\alpha' c}.
\]
Note that rescaling of the integral region doesn't change the leading order of the integral
because $\epsilon^{\alpha'}$ and $(\epsilon^{1/c_i})^{\alpha'}$ both behave as $1+O(\alpha')$
when $\alpha'\to 0$. Thus, it only introduces a factor $\prod_{i\in S}c_i$ 
for the integral. 
Due to this new operation, we don't call it a winning strategy, but it can also be applied to other
similar problems, \textit{e.g.} sector decomposition.

Let's first introduce the matrix language for future use.
For the polynomial $p=\sum_I a_I x^{n^I}$, we introduce the matrix 
\[
	(n_{i}^{I})=
\begin{pmatrix}
	n_1^1 & n_1^2 & \cdots & n_1^N\\
	n_2^1 & n_2^2 & \cdots & n_1^N\\
	\vdots & \vdots & \ddots & \vdots\\
	n_D^1 & n_D^2 & \cdots & n_D^N
\end{pmatrix},
\]
then variable substitutions used by blowing up
\[
	x_j\to x_ix_j \quad \text{for $j\neq i$ and $j\in S$}
\] 
is just replacing the $i$-th row with $\sum_{j\in S} n_j^I$.

\begin{defi}
If a matrix $A=(A^I_i)_{1\leq i\leq D,1\leq I\leq N}$ 
can be obtained from another matrix $B$ by joining/deleting constant row vectors%
\footnote{Here a constant vector is a vector with identical components.},
joining/deleting column vectors $w$ which are in the cone $A^I+\mathbb R_+^D$ for 
some $I$ or adding/subtracting a matrix $C$ whose row vectors are all constant vectors,
then we say that $A$ and $B$ are equivalent, denoted by $A\cong B$.
It's an equivalence relation. In a given equivalence class, we call matrices with minimal number of rows and columns 
reduced matrices.
\end{defi}

For example, 
\[
\begin{pmatrix}
	1&2&3\\
	3&2&4\\
	2&2&3
\end{pmatrix}\cong
\begin{pmatrix}
	0&1&2\\
	1&0&2\\
	0&0&1
\end{pmatrix}\cong
\begin{pmatrix}
	0&1\\
	1&0\\
	0&0
\end{pmatrix}\cong
\begin{pmatrix}
	0&1\\
	1&0
\end{pmatrix} 
\]
If two matrices $(n^I_i)$ and $(m^J_k)$ are equivalent, then there exists a vector $v$ such that 
$\sum_I a_I x^{n^I}\approx x^v\sum_J a_J x^{m^J}$, so we only need to consider blow-up prescriptions for one of equivalent matrices.

\vspace{2ex}

Our algorithm is based on the following observations:
\begin{enumerate}
	\item[0.](Notation). The row vectors are $N$-vector and the column vectors are $D$-vector by default. We mainly consider the row vectors whose index are capital letters, \textit{e.g.} $I$, $J$, .... 
	The coordinate variables of the row vectors are denoted  $\{z^I\}$.
	For a given $i$, $n_i$ is the vector $(n_i^1,\dots,n_i^N)$.
	
	\item If $(n^I_i)$ is equivalent to a $1\times N$ or $D\times 1$ matrix, player $\mathcal{P}_{1}$ wins. 
	Therefore, if we can choose a set $S$ for a given matrix such that $N$ or $D$ 
	decreases strictly in generated matrices, then it's a winning strategy.

	\item If there's exist a set of positive integers $\{c_i\,:\,i\in S\}$ and a 
	positive integer $k$ such that
	\begin{equation}\label{linearprog}
		\sum_{i\in S} c_i n_i = (k,k,\dots,k),
	\end{equation}	
	geometrically this means that the constant vector $(1,\dots,1)$ is in the cone spanned by row vectors $\{n_i\,:\,i\in S\}$,  then by taking variable substitutions $x_{i}\to x_{i}^{1/c_{i}}$  for $i\in S$ and performing the blowing up, 
	the polynomial $p$ becomes
	\[
		p\mapsto x_i^k q
	\]
	where $q$ is $x_{i}$-independent. 
	It's equivalent to the transformation $X_i\mapsto X_i-ck$ and $p\mapsto q$ 
	in the integral. The new generated matrix for each $i\in S$ is just the matrix
	obtained by deleting the $i$-th row of the old matrix. 
	Therefore, we can reduce the number of variables of the polynomial or equivalently 
	the number of \emph{rows} of the matrix in this case.
	
	\item In the viewpoint of cones, 
	a blow-up is nearly a subdivision of the cone.
	If a blow-up produces a vector $\sum_{i\in S} c_i n_i$ in the cone 
	or on the boundary of the cone, 
	then the $j$-th generated cone for $j\in S$ is spanned by 
	vectors $\{n_i\,:\,i\neq j\}$ and the vector $\sum_{i\in S} c_i n_i$,
	which is a proper sub-cone of the original cone. 
	The union of these cones is the original cone,
	but they usually intersect with each other, so it's not a subdivision. 

	For future convenience, if the blow-up produces a vector $v$, we call
	it \textit{the blow-up along $v$}.
	Note that it may be confused with the similar terminology used 
	in the usual mathematical context.

	For example, consider a cone with $D=4$ and $N=3$
	\begin{center}
		\begin{tikzpicture}[scale=0.75]
		\draw[fill=gray!30] (1.8,2.4) node[fill=red,circle,inner sep=1pt] {} 
			-- (3.2,0.8) node[fill=red,circle,inner sep=1pt] {}
			-- (2,-0.5)  node[fill=red,circle,inner sep=1pt] {}
			-- (1.2,0.8) node[fill=blue,circle,inner sep=1pt] {}
			-- cycle;
		\draw (2.8,3.2) -- (-2,-0.5);
		\draw (-2,-0.5) -- (3.5,-0.5);
		\draw[dashed] (-2,-0.5) -- (4,1);
		\draw (-2,-0.5) -- (4,2);
		\draw (3.6,1.3) -- (-2,-0.5);
		\node[fill=green!70!black,circle,inner sep=1pt] at (2.4,0.9) {};
		\end{tikzpicture}
	\end{center}
	We can use the gray polygon, the intersection of this cone with a hyperplain,
	to represent it, and blow up along the green vector which is the positive
	linear combination of three red vectors, then we can get three new subcones
	as shown in the following diagram.
	\begin{center}
		\begin{tikzpicture}[scale=0.75,baseline={([yshift=-.5ex]current bounding box.center)}]
		\draw[fill=gray!30] (2.5,0.2) node[fill=red,circle,inner sep=1pt] {}
			-- (1.6,-1.6) node[fill=blue,circle,inner sep=1pt] {}
			-- (3.2,-2.8) node[fill=red,circle,inner sep=1pt] {}
			-- (4,-1.5) node[fill=red,circle,inner sep=1pt] {}
			-- cycle;
		\node[fill=green!70!black,circle,inner sep=1pt] at (3.1,-1.3) {};
		\end{tikzpicture}
		$\longmapsto$
		\begin{tikzpicture}[scale=0.75,baseline={([yshift=-.5ex]current bounding box.center)}]
		\draw[gray] (2.5,0.2) node[fill=red,circle,inner sep=1pt] {}
			-- (1.6,-1.6) node[fill=blue,circle,inner sep=1pt] {}
			-- (3.2,-2.8) node[fill=red,circle,inner sep=1pt] {}
			-- (4,-1.5) node[fill=red,circle,inner sep=1pt] {}
			-- cycle;
		\draw[fill=gray!30] (2.5,0.2) -- (3.1,-1.3) -- (3.2,-2.8) -- (1.6,-1.6) --cycle;
		\node[fill=green!70!black,circle,inner sep=1pt] at (3.1,-1.3) {};
		\end{tikzpicture}
		\begin{tikzpicture}[scale=0.75,baseline={([yshift=-.5ex]current bounding box.center)}]
		\draw[gray] (2.5,0.2) node[fill=red,circle,inner sep=1pt] {}
			-- (1.6,-1.6) node[fill=blue,circle,inner sep=1pt] {}
			-- (3.2,-2.8) node[fill=red,circle,inner sep=1pt] {}
			-- (4,-1.5) node[fill=red,circle,inner sep=1pt] {}
			-- cycle;
		\draw[fill=gray!30] (3.1,-1.3) -- (4,-1.5) -- (3.2,-2.8) -- (1.6,-1.6) --cycle;
		\node[fill=green!70!black,circle,inner sep=1pt] at (3.1,-1.3) {};
		\end{tikzpicture}
		\begin{tikzpicture}[scale=0.75,baseline={([yshift=-.5ex]current bounding box.center)}]
		\draw[gray] (2.5,0.2) node[fill=red,circle,inner sep=1pt] {}
			-- (1.6,-1.6) node[fill=blue,circle,inner sep=1pt] {}
			-- (3.2,-2.8) node[fill=red,circle,inner sep=1pt] {}
			-- (4,-1.5) node[fill=red,circle,inner sep=1pt] {}
			-- cycle;
		\draw[fill=gray!30] (2.5,0.2) -- (4,-1.5) -- (1.6,-1.6) -- (1.6,-1.6) --cycle;
		\node[fill=green!70!black,circle,inner sep=1pt] at (3.1,-1.3) {};
		\end{tikzpicture}
	\end{center}



	
	The bonus of this viewpoint is that we can see the information of many rounds 
	in only one picture.

	\item If $n^J_i \geq n^I_i$ for all $i$, we can drop column vector $n^J$. Geometrically, 
	it means that the cone spanned by $\{n_i\}$ is in the semi-space defined by $z^J\geq z^I$.
	Thus let's define $H^{IJ}$ as the semi-space defined by $z^I\leq z^J$ 
	in the space of column vectors for future use. 
	Therefore, if we can `divide' the cone by blow-ups into small cones such that 
	each of them is totally contained in some semi-spaces $\{H^{IJ}\}$, then
	the numbers of columns of generated matrices decrease strictly.

	The possible obstacle to do this is that the intersection of 
	all hyperplains $h^{IJ}=\{z\,:\, z^I=z^J\}$, 
	or equivalently the vector $(1,1,\dots,1)$, 
	is contained in the interior of the cone.
	In this case, no matter how blow-ups `divide' the cone, 
	there always exists a cone containing
	$(1,1,\dots,1)$ so that it is not contained in any semi-space $H^{IJ}$.
	However, in this case, we can reduce the number of rows from the observation
	2.

\item 
	One possible way to `divide' the cone is to reduce the number 
	of outside vertices.
	Before giving the explicit definition of outside or inside, 
	let's first consider a $N=3$ example to show this idea.
	\begin{center}
	\begin{tikzpicture}
		\draw (2,-1.5) -- (1.5,-2.5) -- (2,-4) -- (4,-3.5) -- (4.5,-2) -- (3,-1) -- cycle;
		\fill[gray!30] (2,-1.5) -- (1.5,-2.5) -- (2,-4) -- (4,-3.5) -- (4.5,-2) -- (3,-1) -- cycle;
		\draw[thick,densely dotted] (1,-1) -- (4.5,-6);
		\draw[thick,densely dotted] (2.5,-1) -- (3.5,-6);
		\draw[thick,densely dotted] (4,-1) -- (2.5,-6);
		\node[circle,inner sep=1.5pt,fill=blue] at (2,-4) {};
		\node[circle,inner sep=1.5pt,fill=blue] at (4.5,-2) {};
		\node[circle,inner sep=1.5pt,fill=blue] at (1.5,-2.5) {};
		\node[circle,inner sep=1.5pt,fill=blue] at (4,-3.5) {};
		\node[circle,inner sep=1.5pt,fill=red] at (2,-1.5) {};
		\node[circle,inner sep=1.5pt,fill=red] at (3,-1) {};
		\draw[green!60!black] (0.6,-4.5) -- (5.6,-3.5);
		\node at (2.1,-6.2) {$z^{1}{=}z^{2}$};
		\node at (3.5,-6.2) {$z^{2}{=}z^{3}$};
		\node at (4.9,-6.2) {$z^{1}{=}z^{3}$};
	\end{tikzpicture}
	\end{center}
In above diagram, the outside vertices are labeled by blue points, and inside vertices are
labeled by red points. If we blow up the cone along a inside vector which is 
the linear combination of outside vertices, then the number of outside
vertices of each subcone is reduced by $1$. Therefore, after finite such operation,
any subcone is contained in some semi-spaces defined by outside hyperplains.

\item
	In higher dimensional space, hyperplains $h^{IJ}$ divide 
	the whole space $\mathbb R_+^N$ into many small cones which 
	can be labeled by a permutation of $(1,\dots,N)$. 
	The region labeled by $\eta$ is given by the inequality 
	$0\leq z^{\eta_1}\leq \cdots\leq z^{\eta_N}$. 
	We only need to figure out which regions are inside or outside 
	after giving a cone, \textit{then inside (outside) vertices are vertices 
	in the interior of inside (outside) regions.}
	
	For a convex given cone not containing $(1,1,\dots,1)$, 
	we can find a hyperplain $H:\hat{n}\cdot z=0$ crossing 
	$(1,1,\dots,1)$ (\textit{e.g.} the green line in the above diagram) 
	such that the cone is contained in one side of this hyperplain.
	In other words, we are looking for a vector $\hat{n}\in \mathbb R^N$ 
	such that
	\[
		\sum_{I=1}^N \hat{n}^I=0\quad \text{and}\quad \hat{n}\cdot
		n_i=\sum_{I=1}^N\hat{n}^In^I_i\geq 0 \text{ for all $i$}.
	\]
	Note that, in $N=3$ case, a outside region can cross the hyperplain $H$ 
	but a inside region cannot. We just generalize this criterion to higher dimension.
	
	Equivalently, a region is inside (outside) if and only if $\hat{n}$ 
	is (not) contained in its dual cone spanned by normal vectors of 
	surrounding hyperplains $\{h^{IJ}\}$. Precisely, 
	for the region $R$ label by a permutation $\eta$,
	its dual cone is spanned by
	\[
		\{\mathbf{b}^I_\eta=\mathbf{e}_{\eta_{I+1}}-\mathbf{e}_{\eta_{I}}
		\,:\, 1\leq I\leq N-1\},
	\]
	where $\mathbf{e}_i\in \mathbb R^N$ is the vector whose $i$-th element 
	is $1$ and the other elements are all zero, then
	the region $R$ is inside if and only if there exist nonnegative numbers
	$\{c_I\}$ such that $\hat{n}=\sum_I c_I\mathbf{b}^I_\eta$.
\end{enumerate}
Now our algorithm (for one polynomial) is simple: For a given cone $(n^I_i)$,
\begin{enumerate}
	\item[(0).] Define a set of matrices $\mathcal M$ and initialize it 
		to $\{(n^I_i)\}$. Note that we may add matrices with different 
		sizes to $\mathcal M$ in the algorithm, so the dimension of 
		$(1,1,\dots,1)$ in the context depends on the matrix.
	\item[(1).] Replace any matrix by its reduced matrix. If all matrices
		in the set $\mathcal M$ have only one column or row, 
		the algorithm stops.
	\item[(2).] 
		Look for cones containing $(1,1,\dots,1)$ in $\mathcal M$.
		If there's no such cone, goto step (3).
		Otherwise, blow up such matrices along $(1,1,\dots,1)$, 
		add all generated matrix into the set $\mathcal M$ and goto step (1).
	\item[(3).] 
		For a matrix $C$ with more than one column in $\mathcal M$, choose two outside vertices $n_i$ and $n_j$ in different regions.
		Since there exists a hyperplane $h^{AB}$ separating them, 
		we can find positive integers $p$ and $q$ such that
		$pn_i+qn_j$ is on the hyperplane $h^{AB}$. 
		Blow up this cone $C$ along this vector so that the number of 
		outside vertices is reduced by one, add generated matrices into 
		the set $\mathcal M$ and goto step (1).
\end{enumerate}
This algorithm always terminates in finite steps.

Finally we use a simple example to end this section. Consider the matrix
\[
	M=\left(
		\begin{array}{ccc}
			n_1 \\
			n_2 \\
			n_3
		\end{array}
	\right)
	=\left(
		\begin{array}{ccc}
			0 & 3 & 1 \\
			2 & 0 & 1 \\
			1 & 2 & 0
		\end{array}
	\right),
\]
we can represent the cone by the projection of its intersection with 
the hyperplain $z^1+z^2+z^3=12$ on the $z^1$-$z^2$ plane.
\begin{center}
\begin{tikzpicture}
\fill[opacity=0.2,gray] (0,4.5) -- (4,0) -- (2,4) -- cycle ;
\draw[thick] (0,4.5) -- (4,0) -- (2,4) -- cycle ;
\draw[->] (-0.1,0) -- (5,0) node[below] {$z^1$};
\draw[->] (0,-0.1) -- (0,5) node[left] {$z^2$};
\draw[densely dotted] (0,0) -- (4,4) node[right] {$z^1=z^2$};
\draw[densely dotted] (4.2,0.9)  node[right] {$z^2=z^3$} -- (0.6,2.7);
\draw[densely dotted] (0.7,4.6)  -- (2.7,0.6)  node[below] {$z^1=z^3$};
\draw[thick,dashed] (2,4) -- (1.7,2.6) node[circle, fill=red, inner sep=1pt] {};
\node at (2.2,4.3) {$n_3$};
\node at (4.1,-0.3) {$n_2$};
\node at (-0.3,4.5) {$n_1$};
\end{tikzpicture}
\end{center}
The intersection of three dotted line is the vector $(4,4,4)$, 
so the cone doesn't contain $(1,1,1)$ and we can directly goto step (3).
The outside vertices are $n_1$ and $n_2$, so we first find a inside red point 
as the intersection of the hyperplain 
$z^1=z^3$ and the facet spanned by $\{n_1,n_2\}$ of the cone, so in fact the red point represents the vector $n_1+n_2$, which tells us that we should blow up $x_1$ and $x_2$. Therefore, the cone is decomposed into two new cones 
\[
	M\mapsto \biggl\{
		\begin{pmatrix}
			0&3&1\\
			2&3&2\\
			1&2&0
		\end{pmatrix},
		\begin{pmatrix}
			2&3&2\\
			2&0&1\\
			1&2&0
		\end{pmatrix}
	\biggr\}
\] 
by blowing up, and these matrices can be reduced to
\[
	M\mapsto \biggl\{
		\begin{pmatrix}
			0&1\\
			2&2\\
			1&0
		\end{pmatrix}\cong
		\begin{pmatrix}
			0&1\\
			1&0
		\end{pmatrix},
		\begin{pmatrix}
			3&2\\
			0&1\\
			2&0
		\end{pmatrix}
	\biggr\}
\]
because the first one is contained in the semi-space $z^3\leq z^2$
and the second one is contained in the semi-space $z^3\leq z^1$.
It's easy to see that all new cones contain $(1,1)$, so we goto step (1) 
and get
\begin{align*}
	M&\mapsto 
	\biggl\{
		\begin{pmatrix}
			0&1\\
		\end{pmatrix},
		\begin{pmatrix}
			1&0\\
		\end{pmatrix},
		\begin{pmatrix}
			0&1\\
			2&0
		\end{pmatrix},
		\begin{pmatrix}
			3&2\\
			2&0
		\end{pmatrix}
	\biggr\}\\
	 &\cong
	 \biggl\{
		 \begin{pmatrix}
			 0\\
		 \end{pmatrix},
		 \begin{pmatrix}
			 0\\
		 \end{pmatrix},
		 \begin{pmatrix}
			 0&1\\
			 2&0
		 \end{pmatrix},
		 \begin{pmatrix}
			 2\\
			 0
		 \end{pmatrix}
	 \biggr\}\\
	 &\mapsto
	 \biggl\{
		 \begin{pmatrix}
			 0\\
		 \end{pmatrix},
		 \begin{pmatrix}
			 0\\
		 \end{pmatrix},
		 \begin{pmatrix}
			 2&0
		 \end{pmatrix},
		 \begin{pmatrix}
			 0&2\\
		 \end{pmatrix},
		 \begin{pmatrix}
			 2\\
			 0
		 \end{pmatrix}
	 \biggr\}
\end{align*}
Now the algorithm stops. If we go back to the integral of the polynomial $p=y^2z+x^3z^2+xy$
corresponding to $M$, we should carefully add factors $c_i$
corresponding to the change of variables $x_i\mapsto x_i^{1/c_i}$ and we will get $5$ terms 
in above example.
\begin{align*}
	\int_{[0,\epsilon]^3}\frac{\dif x}{x}&\frac{\dif y}{y}\frac{\dif z}{z}
	x^{\alpha' X}y^{\alpha' Y}z^{\alpha' Z}(y^2z+x^3z^2+xy)^{-\alpha' c}\\
	=\,&\frac{1}{{\alpha'}^3}\biggl(
	\frac{1}{X (-2 c+X+Y) (-c+X+Z)}+\frac{1}{Z (-2 c+X+Y) (-c+X+Z)}\\
	&+\frac{1}{Z (-2 c+X+Y) (-3 c+X+2 Y)}+
	\frac{2}{Z (-3 c+X+2 Y) (-2 c+2 Y+Z)}\\
	&+\frac{2}{2 Y (-3 c+X+2 Y) (-2 c+2 Y+Z)}
	\biggr) +O({\alpha'}^{-2}).
\end{align*}

\section{Conclusions and Outlook}

In this article, we have shown how to obtain the leading order contribution of stringy integrals by a blow-up algorithm. These integrals are generations of tree-level scattering of open strings and provide natural $\alpha'$-deformed canonical forms for general polytopes. Interestingly, the algorithm is equivalent to a winning strategy for a simplified version of Hironaka's polyhedra game since more moves can be taken in our case.

All information about the leading order of such integrals is contained in the Minkowski sum $N_{P}$ of Newton polytopes of the regulating polynomials. In this sense, the blow-up method gives a way to reconstruct the polytope $N_{P}$ from its vertices. However, as we saw in section 3, spurious poles and hence spurious vertices are produced in the process of blow-ups. Additional efforts are still needed to recognize the real poles and vertices, especially when the dimensionality increases (curse of dimensionality) and the regulating polynomials get more and more complicated. The result obtained by blowing up is correct but redundant, thus a interesting question is how the polytope $N_{P}$ emerges from this (usually tedious) result.

Another way to reconstruct $N_{P}$ is, along with the opposite direction, to find the facets of $N_{P}$ by using the scattering-equation map. Where the aim is to find all directions such that $\mathbf{X}$ approaches facets of $N_{P}$ as $\mathbf{x}$ approaches boundaries of $\mathbb{R}^{D}$ along with these directions~\cite{He:2020ray}. It would be interesting to find any relations between the reconstruction from the bottom up, the blow-up, and the reconstructing from the top down, the scattering-equation map.

As we saw in section 2, lots of terms have been dropped during the process of blow-ups, while higher order contributions of such integrals with respect to $\alpha'$ certainly depend on the details of regulating polynomials. One way to obtain the higher order contribution is to expand the integrand with respect to $\alpha'$ then integration, the obstacle to this expansion is singularities in poles. The blow-up procedure provide a way to remove this obstacle: each integration region produced by blow-ups only meets singularities of the canonical form at one vertex, then a subtraction can be easily made such that the integrand have no divergence in the integration region.

The algorithm introduced in section 4 is, in some sense, the byproduct of the geometric viewpoint, and there is still much room to improve. Some possible improvements could happen in the procedure (3) of the algorithm. 
For example, we haven't figured out what's the most efficient way to choose 
two outside vertices to blow up, and it should be more efficient to blow up along vectors in the inside region than
vectors on hyperplanes $\{h^{IJ}\}$. One should even invent more efficient new methods 
to realize the procedure (3), \textit{i.e.} reduce the number of columns of matrices, 
from this viewpoint. What's more, it's believed that this new viewpoint could bring
us a new winning procedure of the original Hironaka's polyhedra game. 
We leave these for the future.

Since the situation here is similar with the sector decomposition,
we expect that this kind of geometric viewpoint will give some insights of the calculation of Feynman diagrams in which some efforts have been made (see \textit{e.g.} \cite{Kaneko:2009qx}). On the other hand, another object for Feynman diagrams, Hepp's bound, is recently found to have a closed relation to the polytope geometry~\cite{Panzer:2019yxl}. 
It would be fascinating to explore further relations between the leading order contribution of the stringy canonical forms and Feynman diagrams from this geometric viewpoint.

\section*{Acknowledgement}
The original idea for this work came from the study of positive geometries, stringy canonical forms and integrals related to cluster associahedra, by Nima Arkani-Hamed, Song He and Thomas Lam, to whom we are grateful for suggesting the project, sharing ideas and many valuable inputs.  C.Z. thanks the Institute for Advanced Study, Princeton for hospitality during the start stage of the work. We also thank Qinglin Yang for useful discussions and collaborations in an early stage of the project.

\bibliographystyle{unsrt}
\bibliography{main}

\begin{thebibliography}{10}

\bibitem{Arkani-Hamed:2019mrd}
Nima Arkani-Hamed, Song He, and Thomas Lam.
\newblock {Stringy Canonical Forms}.
\newblock 2019.

\bibitem{hironaka1967}
Heisuke Hironaka.
\newblock Characteristic polyhedra of singularities.
\newblock {\em J. Math. Kyoto Univ.}, 7(3):251--293, 1967.

\bibitem{Binoth:2000ps}
T.~Binoth and G.~Heinrich.
\newblock {An automatized algorithm to compute infrared divergent multiloop
  integrals}.
\newblock {\em Nucl. Phys.}, B585:741--759, 2000.

\bibitem{Binoth:2003ak}
T.~Binoth and G.~Heinrich.
\newblock {Numerical evaluation of multiloop integrals by sector
  decomposition}.
\newblock {\em Nucl. Phys.}, B680:375--388, 2004.

\bibitem{Heinrich:2008si}
Gudrun Heinrich.
\newblock {Sector Decomposition}.
\newblock {\em Int. J. Mod. Phys.}, A23:1457--1486, 2008.

\bibitem{shafarevich1994basic}
Igor'~Rostislavovich Shafarevich.
\newblock {\em Basic algebraic geometry}, volume~1.
\newblock Springer, 1994.

\bibitem{Arkani-Hamed:2017tmz}
Nima Arkani-Hamed, Yuntao Bai, and Thomas Lam.
\newblock {Positive Geometries and Canonical Forms}.
\newblock {\em JHEP}, 11:039, 2017.

\bibitem{spivakovsky1983solution}
Mark Spivakovsky.
\newblock A solution to hironaka’s polyhedra game.
\newblock In {\em Arithmetic and geometry}, pages 419--432. Springer, 1983.

\bibitem{zeillinger2005polyederspiele}
Dominik Zeillinger.
\newblock {\em Polyederspiele und Aufl{\"o}sen von Singularit{\"a}ten}.
\newblock na, 2005.

\bibitem{zeillinger2006short}
Dominik Zeillinger.
\newblock A short solution to hironaka's polyhedra game.
\newblock {\em ENSEIGNEMENT MATHEMATIQUE}, 52(1/2):143, 2006.

\bibitem{hauser2003hironaka}
Herwig Hauser.
\newblock The hironaka theorem on resolution of singularities (or: A proof we
  always wanted to understand).
\newblock {\em Bulletin of the American Mathematical Society}, 40(3):323--403,
  2003.

\bibitem{Smirnov:2008py}
A.~V. Smirnov and M.~N. Tentyukov.
\newblock {Feynman Integral Evaluation by a Sector decomposiTion Approach
  (FIESTA)}.
\newblock {\em Comput. Phys. Commun.}, 180:735--746, 2009.

\bibitem{Ueda:2009xx}
Takahiro Ueda and Junpei Fujimoto.
\newblock {New implementation of the sector decomposition on FORM}.
\newblock {\em PoS}, ACAT08:120, 2008.

\bibitem{Arkani-Hamed:2019plo}
Nima Arkani-Hamed, Song He, Thomas Lam, and Hugh Thomas.
\newblock {Binary Geometries, Generalized Particles and Strings, and Cluster
  Algebras}.
\newblock 2019.

\bibitem{Mafra:2016mcc}
Carlos~R. Mafra and Oliver Schlotterer.
\newblock {Non-abelian $Z$-theory: Berends-Giele recursion for the
  $\alpha'$-expansion of disk integrals}.
\newblock {\em JHEP}, 01:031, 2017.

\bibitem{fomin2007cluster}
Sergey Fomin and Andrei Zelevinsky.
\newblock Cluster algebras iv: coefficients.
\newblock {\em Compositio Mathematica}, 143(1):112--164, 2007.

\bibitem{Li:2018mnq}
Zhenjie Li and Chi Zhang.
\newblock {Moduli Space of Paired Punctures, Cyclohedra and Particle Pairs on a
  Circle}.
\newblock {\em JHEP}, 05:029, 2019.

\bibitem{He:2020ray}
Song He, Lecheng Ren, and Yong Zhang.
\newblock {Notes on polytopes, amplitudes and boundary configurations for
  Grassmannian string integrals}.
\newblock 2020.

\bibitem{Kaneko:2009qx}
Toshiaki Kaneko and Takahiro Ueda.
\newblock {A Geometric method of sector decomposition}.
\newblock {\em Comput. Phys. Commun.}, 181:1352--1361, 2010.

\bibitem{Panzer:2019yxl}
Erik Panzer.
\newblock {Hepp's bound for Feynman graphs and matroids}.
\newblock 2019.

\end{thebibliography}

\end{document}